\documentclass[aps,preprint]{revtex4}
\usepackage{amsmath}
\usepackage[english]{babel}
\usepackage{amssymb}
\usepackage{graphicx}
\usepackage{epsfig}
\usepackage{bm}
\usepackage{hyperref}
\usepackage{color}

\begin{document}

\title{\sc\Large{Magnetic field driven enhancement of the weak decay width of charged pions} \vspace*{.5cm}}

\author{M. Coppola$^{a,b}$, D. Gomez Dumm$^{c}$, S. Noguera$^{d}$ and N.N.\ Scoccola$^{a,b}$ \vspace*{.3cm}}

\affiliation{$^{a}$ CONICET, Rivadavia 1917, 1033 Buenos Aires, Argentina}
\affiliation{$^{b}$ Physics Department, Comisi\'{o}n Nacional de Energ\'{\i}a At\'{o}mica, }
\affiliation{Av.\ Libertador 8250, 1429 Buenos Aires, Argentina}
\affiliation{$^{c}$ IFLP, CONICET $-$ Departamento de F\'{\i}sica, Fac.\ de Cs.\ Exactas,
Universidad Nacional de La Plata, C.C. 67, 1900 La Plata, Argentina}
\affiliation{$^{d}$ Departamento de F\'{\i}sica Te\'{o}rica and IFIC, Centro Mixto
Universidad de Valencia-CSIC, E-46100 Burjassot (Valencia), Spain \vspace*{2cm}}

\begin{abstract}
We study the effect of a uniform magnetic field $\vec B$ on the
decays $\pi^-\to l^-\,\bar{\nu}_l$, where $l^-\,=e^-,\,\mu^-$, carrying out a
general analysis that includes four $\pi^-$ decay constants. Taking the
values of these constants from a chiral effective Nambu-Jona--Lasinio (NJL)
model, it is seen that the total decay rate gets strongly increased with
respect to the $B=0$ case, with an enhancement factor ranging from $\sim 10$
for $eB=0.1\,\mathrm{GeV}^2$ up to $\sim 10^3$ for $eB=1\,\mathrm{GeV}^2$.
The ratio between electronic and muonic decays gets also enhanced, reaching
a value of about $1:2$ for $eB=1\,\mathrm{GeV}^2$. In addition, we find that
for large $B$ the angular distribution of outgoing antineutrinos shows a
significant suppression in the direction of the magnetic field.
\end{abstract}

\maketitle

\section{Introduction}
\label{sec:intro}

The effect of intense magnetic fields on the properties of strongly
interacting matter has gained significant interest in recent
years~\cite{Kharzeev:2012ph,Andersen:2014xxa,Miransky:2015ava}. This is
mostly motivated by the realization that strong magnetic fields might play
an important role in the study of the early Universe~\cite{Grasso:2000wj},
in the analysis of high energy non-central heavy ion
collisions~\cite{Kharzeev,Skokov,Voronyuk}, and in the description of
compact stellar objects like the magnetars~\cite{Duncan,Kouveliotou}. It is
well known that magnetic fields also induce interesting phenomena such as
the enhancement of the QCD vacuum (the so-called ``magnetic
catalysis'')~\cite{Gusynin:1994re} and the decrease of critical temperatures
for chiral restoration and deconfinement QCD
transitions~\cite{Bali:2011qj,Bali:2012zg}. In this work we concentrate on
the effect of a magnetic field $\vec B$ on the weak pion-to-lepton decays
$\pi^-\to l^-\,\bar\nu_l$, where $l^-\,=e^-,\,\mu^-$. In fact, the study of weak
decays of hadrons in the presence of strong electromagnetic fields has a
rather long history (see
e.g.~refs.~\cite{Nikishov:1964zza,Nikishov:1964zz,Matese:1969zz,FassioCanuto:1970wk}).
In most of the existing calculations of these decay rates, however, the
effect of the external field on the internal structure of the participating
particles has not been taken into account. In the case of charged pions,
only recently such an effect has been analyzed in the context of chiral
perturbation theory~\cite{Andersen:2012zc} and effective chiral
models~\cite{Simonov:2015xta, Liu:2018zag, Coppola:2018vkw}, as well as
through lattice QCD (LQCD) calculations~\cite{Bali:2018sey}. In
ref.~\cite{Bali:2018sey} it is noted that the existence of the background
field opens the possibility of a nonzero pion-to-vacuum transition via the
vector piece of the hadronic current, implying the existence of a further
form factor in addition to the pion decay constant $f_\pi$ (which arises
from the axial vector piece). Taking into account this new decay constant
and using some approximations for the dynamics of the participating
particles, the authors of ref.~\cite{Bali:2018sey} obtain an expression for
the $\pi^-$ decay width in the presence of the external field. In
particular, it is claimed that for $eB\sim 0.3$~GeV$^2$, $e$ being the
proton charge, the decay rate of charged pions into muons could be enhanced
by a factor of about 50 with respect to its value at $B=0$. Recently, a
more complete analysis of the situation has been presented in
ref.~\cite{Coppola:2018ygv}, where the most general form of the relevant
hadronic matrix elements in the presence of an external uniform magnetic
field was determined. It was found that in general the vector and axial
vector pion-to-vacuum transitions (for the case of charged pions) can be
parametrized through one and three hadronic form factors, respectively.
Taking into account all four decay constants, in ref.~\cite{Coppola:2018ygv}
an expression for the $\pi^-\to l^-\,\bar\nu_l$ decay width that fully takes
into account the effect of the magnetic field on both pion and lepton wave
functions was obtained using the Landau gauge. The same expression was found
in ref.~\cite{Coppola:2019wvh} using the symmetric gauge, explicitly showing
the gauge independence of the result.

The main purpose of this article is to show that, once the above-mentioned
improvements are incorporated, the $\pi^-\to l^-\,\bar\nu_l$ decay rate in the
presence of the magnetic field turns out to be strongly enhanced with
respect to its value for $B=0$. Taking values for the decay constants from
an effective Nambu-Jona-Lasinio (NJL) model, this enhancement is found to
range from $\sim 10$ for $eB=0.1$~GeV$^2$ up to $\sim 10^3$ for
$eB=1$~GeV$^2$. Interestingly, it is found that the ratio between $\pi^-$
partial decay rates into electrons and muons gets also significantly
increased, reaching a value of about 0.5 for $eB=1$~GeV$^2$. In addition, it
is observed that already for $eB\simeq 0.1$~GeV$^2$ the angular distribution
of the outgoing antineutrinos is expected to be highly anisotropic, showing
a significant suppression in the direction of the magnetic field.

The paper is organized as follows. In sec.~\ref{sec:decay} we present a
general theoretical analysis of the $\pi^-\to l^-\,\bar\nu_l$ decay width in
the presence of the external field. This includes a comparison with the
$B=0$ case and a discussion on the lack of the helicity suppression
mechanism. In sec.~\ref{sec:results}, numerical estimations are given in the
framework of the NJL model. Finally, in sec.~\ref{sec:summary} we summarize
our research and provide some conclusions. We also include two
appendices. In appendix~\ref{appA} we present a brief discussion on the relation
between gauge invariance and axial rotations, while in appendix~\ref{appB} we
give some expressions for pion and lepton wavefunctions in the presence of
the magnetic field.

\section{\texorpdfstring{$\bm{\pi^- \to l^-\,\bar\nu_l\:}$}{}  decay}
\label{sec:decay}

\subsection{Absence of helicity suppression for nonzero external magnetic field}
\label{subsec:decayH}

As well known, if there is no external magnetic field the decay width
$\Gamma(\pi^- \to l^-\, \bar{\nu}_l)$ in the pion rest frame is given by
\begin{equation}
\Gamma_l^-\,(0) \ =  \ \dfrac{G^2_F\,\cos^2\theta_c}{4\pi}\; \,f^2_{\pi} \ m_\pi\,m_l^2\,
\left(1-\dfrac{m_l^2}{m_\pi^2}\right)^2\ ,
\label{bzero}
\end{equation}
where $G_F$ is the Fermi effective coupling, $\theta_c$ is the Cabibbo
angle, and the value of the decay constant $f_\pi = f(m_\pi^2)\simeq
92.3$~MeV can be obtained from the empirical $\pi^-$ mean lifetime $\tau
\simeq \left(\sum_l \Gamma_l^-\,(0) \right)^{-1} = 2.603 \times
10^{-8}$~s~\cite{Tanabashi:2018oca}. Owing to the $m_l^2$ factor, the total
width is strongly dominated by the muonic decay, for which the branching
ratio reaches about 99.99\%. The reason for this behavior can be easily
understood in terms of ``helicity suppression''. In the pion rest frame, the
outgoing charged lepton and antineutrino have opposite momenta, therefore
the final state has zero orbital angular momentum, and angular momentum
conservation requires both outgoing particles to have opposite spins. Taking
the direction of the momenta as the angular momentum quantization axis, this
implies that the charged lepton $l^-\,$ and the antineutrino $\bar \nu_l$
should have the same helicity. On the other hand, the electroweak current
couples the $\pi^-$ only to right-handed antineutrinos and left-handed
charged leptons. Then, if we assume that neutrinos are massless, the
helicity of the antineutrino will be $+1$. In the limit $m_l \to 0$ the
helicity of the left-handed charged lepton will be $-1$, i.e. opposite to
that of the antineutrino. Since this is in contradiction with the result
above, the decay turns out to be forbidden in that limit.

In the presence of an external uniform magnetic $\vec B$, the above
situation becomes dramatically modified. For definiteness, let us take the
magnetic field to lie along the $z$ axis, $\vec B = (0,0,B)$, with $B>0$. As
in the $B=0$ case, we assume the charged pion to be in its lowest possible
energy state. The latter corresponds to the lowest Landau level (LLL)
$\ell=0$, and the pion $z$ component of the momentum $p_z=0$. It is worth
stressing that, even in this lowest energy state, the decaying pion cannot
be at rest, due to the existence of a nonvanishing zero-point motion. In
fact, the three spacial components of pion momentum are not a good set of
quantum numbers to describe the initial state in this case. Moreover, the
outcomes obtained for $B=0$ from angular momentum conservation do not apply
for nonzero $B$. The analysis of the decay in terms of angular momenta of
the initial and final states is not straightforward, since for nonzero $B$
canonical angular momenta of charged particles turn out to be gauge
dependent quantities, and total mechanical angular momentum is in general
not conserved~\cite{Li,Greenshields,Wakamatsu:2017isl,Coppola:2019wvh}. A
brief discussion on how this can be reconciled with the rotational
invariance of the system is included in appendix~\ref{appA}.

To have a better understanding of the situation, it is interesting to
consider the case in which the magnitude of the magnetic field is large
enough so that the outgoing charged lepton $l^-\,$ can only be in the LLL,
$n=0$ (the validity of this assumption will be discussed below). Considering
the explicit form of the corresponding
spinors~\cite{Coppola:2018ygv,Coppola:2019wvh}, it is not hard to show
(see appendix~\ref{appB}) that in the $m_l \to 0$ limit one has
\begin{equation}
\gamma_5 |l^- (LLL) \rangle \ =
\ \hat Q \cdot \vec \Sigma  |l^- (LLL) \rangle \ =
\ -\, \mbox{sign}(q_z)  \,
|l^- (LLL) \rangle \ , \label{chirality}
\end{equation}
where $\vec Q = \vec q + e\vec A$ is the mechanical linear momentum
operator (a gauge invariant quantity) and $q_z$ is the $z$
component of the momentum of the charged lepton. As expected, the chirality
of the LLL lepton state coincides with its helicity in the massless limit.
Interestingly, in eq.~(\ref{chirality}) only the parallel piece of the
helicity operator contributes. This can be understood by noting that for the
LLL only one polarization state, namely that associated to $\Sigma_z(l^-\,) =
-1$, is allowed (see appendix~\ref{appB}). Being $\Sigma_x$ and
$\Sigma_y$ polarization-changing operators, the action of the sum $Q_x
\Sigma_x + Q_y \Sigma_y$ on the $|l^- (LLL)\rangle$ state has to
vanish in order to ensure that the latter is an helicity eigenstate, as it
should be in the $m_l \to 0$ limit. Let us consider now the outgoing
antineutrino, taking it to be in a state of momentum $\vec k$. Since it has
to be right-handed, the helicity operator satisfies
\begin{equation}
\gamma_5 |\bar \nu_l
\rangle \ = \ \hat k \cdot \vec \Sigma\, |\bar \nu_l \rangle  \ =  \ + \, |\bar \nu_l \rangle \ .
\label{neutrino}
\end{equation}
In this case, however, the transverse piece of the helicity operator
provides in general a nonvanishing contribution. On one hand, there is no
restriction for antineutrino helicity eigenstates to be in general a
combination of the two available possible polarization states,
$\Sigma_z(\bar\nu_l) = \pm 1$~\cite{Coppola:2018ygv,Coppola:2019wvh}. On the
other hand, the antineutrino transverse momentum $\vec k_\perp$ is in
general nonvanishing, since, due to zero-point motion, the wavefunctions of
charged particles in the LLL involve a superposition of various transverse
momenta. Therefore, eq.~(\ref{neutrino}) does not determine the sign of
$k_z$, and nothing forces the outgoing particles to have the same helicity,
in contrast with the $B=0$ case. Thus, no helicity suppression
mechanism is present for nonzero $B$, and, consequently, the $\pi^-\to
l^-\,\bar\nu_l$ decay amplitude does not necessarily vanish in the $m_l \to 0$
limit.

To quantitatively see how important the ``non-helicity suppression''
effect is, one has to analyze in detail the $\pi^- \to l^-\, \bar{\nu}_l$
decay width in the presence of the magnetic field. A model independent
expression for the width has been obtained in
refs.~\cite{Coppola:2018ygv,Coppola:2019wvh}, taking the decaying pion to be
in the LLL, with $p_z = 0$. The main steps leading to this expression are
summarized in the following subsections.

\subsection{Particle states and gauge choice}
\label{subsec:decayS}

The actual calculation of the partial widths $\Gamma(\pi^- \to l^-\,
\bar{\nu}_l)$ for nonzero external magnetic field requires to choose a
specific gauge. We note, however, that the widths are expected to be gauge
independent, as explicitly shown in refs.~\cite{Coppola:2018ygv} and
\cite{Coppola:2019wvh}, where the same result has been obtained considering
the Landau and symmetric gauges, respectively. Here we will retrieve
some of the steps followed for the case of the symmetric gauge, in which one
has axial symmetry and the participating particles can be expressed in terms
of states of well defined angular momentum projection in the direction of
the external field.

For our calculations we adopt the following conventions. For a space-time
coordinate four-vector $x^\mu$ we use the notation $x^\mu = (t, \vec r\,)$,
taking the Minkowski metric $g^{\mu\nu} = \mbox{diag}(1,-1,-1,-1)$. We
assume the presence of a uniform static magnetic field $\vec B$, and
orientate the spatial axes in such a way that $\vec B = B\ \hat z$, with
$B>0$. Owing to axial symmetry, it is convenient to use for $\vec r$
standard cylindrical coordinates $\rho$, $\phi$ and $z$. The vector
potential will be then given by $A^\mu = (0,\vec A)$, with $\vec A = \vec B
\times \vec r/2 = (-B\rho \sin\phi/2, B \rho\cos\phi/2 , 0)$.

As already mentioned, in the presence of an external magnetic field the
three spacial components of momentum are not a good set of quantum numbers
for charged particles. In fact, in the plane perpendicular to
$\vec B$, charged particle states are quantized in Landau levels. For the
symmetric gauge, given our axis choice, one can define a complete basis of states of well defined
energy taking as quantum numbers the $z$ component of the momentum, the
Landau level and the $z$ component of the \textit{canonical} total angular
momentum $\vec j$. For the antineutrino, having zero electric charge, we
take $k_z$, $j_z$ and $k_\perp = \sqrt{k_x^2+k_y^2}$, where $\vec k$ is
the antineutrino linear momentum. The notation used for the quantum numbers
of the $\pi^-$, $l^-$ and $\bar\nu_l$ is summarized in table~\ref{Tprop}.
Here $\ell$, $n$, $\imath$, and $\upsilon$ are non-negative integers,
$\jmath$ is an integer, and $B_e=|e\vec B|$. To this set of quantum numbers one
has to add the polarization $\tau$ ($\tau = 1,2$) of the charged lepton (we
assume the antineutrino to be purely righthanded). Notice that, although it
is not indicated explicitly, the pion mass $m_{\pi^-}$ is a function of the
magnetic field $B$. The explicit form of the $\pi^-$, $l^-$ and $\bar\nu_l$
wavefunctions and spinors in the symmetric gauge is quoted in
appendix~\ref{appB}.

\begin{table}[h]
\begin{center}
\begin{tabular}{| l | c | c | c |}
 \hline
 &  Pion $(\pi^-)$ & Lepton $(l^-\,)$ & \ Antineutrino $(\bar \nu_l)$ \ \\ 
 \hline 
 \ Parallel momentum \ & $p_z$ & $q_z$ & $k_z$  \\
 \ Landau level & $\ell$ & $n$ & --  \\
 \ $j_z$ & $\ell - \imath$ & $n - \upsilon -1/2$ & $\jmath-1/2$ \\
 \ Energy & \ $\sqrt{m_{\pi^-}^2+ (2\ell+1) B_e + p_z^2}\:$ \ & \ $\sqrt{m_l^2 + 2n B_e + q_z^2}\:$ \ &
 $\sqrt{k_\perp^2 + k_z^2}$ \\
 \ Shorthand notation \ & $\breve{p} = (\ell,\imath,p_z)$ & $\breve {q} = (n,\upsilon,q_z)$ &
 $\breve k = (\jmath, k_\perp, k_z)$ \\
 \hline
\end{tabular}
\caption{Notation for particle quantum numbers.}
\label{Tprop}
\end{center}
\end{table}

\subsection{Decay amplitude}
\label{subsec:decayA}

According to the notation introduced in the previous subsection, the
transition matrix element for the $\pi^-\to l^-\,\bar\nu_l$ decay is given by
$\langle\, l^-(\breve q , \tau)\, \bar \nu_l(\breve k, R ) | \mathcal{L}_W |
\pi^-(\breve p)\,\rangle$. As usual, the amplitude can be written in terms
of leptonic and hadronic parts. Taking into account the expressions for the
involved fields quoted in appendix~\ref{appB} (for more details, see also
ref.~\cite{Coppola:2019wvh}) one gets
\begin{align}
\langle\, l^-(\breve q , \tau)\, \bar \nu_l(\breve k, R ) | \mathcal{L}_W |
\pi^-(\breve p)\,\rangle = &  - \dfrac{G_F}{\sqrt{2}}\, \cos \theta_c \, \times
\nonumber \\
&  \times \int d^4x \, H^{\mu}_L(x,\breve p)
\ \bar  U^-_l(x,\breve {q}, \tau) \, \gamma_\mu \, (1-\gamma_5) \,
V_{\nu_l}(x,\breve k, R)\ , \label{transition}
\end{align}
where $H^{\mu}_L(x,\breve p)$ stands for the matrix element of the hadronic current,
\begin{equation}
H^{\mu}_{L}(x,\breve p) \ = \
H_V^{\mu} (x,\breve p) - H_A^{\mu} (x,\breve p) \ = \
\langle 0| \bar \psi_u(x)\, \gamma^\mu (1-\gamma_5)\, \psi_d(x)| \pi^-(\breve p) \rangle\ .
\label{hadronicme}
\end{equation}

The matrix element in eq.~(\ref{hadronicme}) involves strong interactions in
a low energy regime and cannot be treated perturbatively. Instead, it can be
parameterized in terms of decay form factors taking into account the Lorentz
structure and the symmetries of the theory. As it is well known, in the
absence of external fields the amplitude can be written in terms of a single
form factor, namely, the pion decay constant $f_\pi$. In that case, owing to
parity symmetry, only the axial-vector piece $H_A^{\mu}$ can be nonzero.
However, when a static external electromagnetic field is present, several
independent tensor structures are allowed and four independent form factors
can be defined. Three of them correspond to the axial-vector and one to the
vector piece of the hadronic current. Following ref.~\cite{Coppola:2018ygv},
the hadronic matrix element in eq.~(\ref{hadronicme}) can be parameterized
as
\begin{eqnarray}
H^{\mu}_{L}(x,\breve p) & = &
\left[ \epsilon^{\mu\nu\alpha\beta} F_{\nu\alpha}{\cal D}_{\beta} \, \dfrac{f_{\pi^-}^{(V)}}{2B} -
{\cal D}^{\mu} \, f_{\pi^-}^{(A1)} \, +
i\, F^{\mu\nu}{\cal D}_{\nu}\, \dfrac{f_{\pi^-}^{(A2)}}{B}\, -
F^{\mu\nu}F_{\nu\alpha}{\cal D}^{\alpha}\, \dfrac{f_{\pi^-}^{(A3)}}{B^2} \
\right]\,\times  \nonumber\\[.15cm]
& &  \sqrt{2} \, \langle 0 | \phi_{\pi^-} (x) | \pi^-(\breve p) \rangle \ ,
\label{hadamp}
\end{eqnarray}
where $F^{\mu\nu}$ is the electromagnetic field tensor, and $\mathcal{D}^\mu
= \partial^\mu - ieA^\mu$. It can be seen that the discrete symmetries of
the interaction Lagrangian restrict all four form factors to be
real~\cite{Coppola:2018ygv}. In the symmetric gauge, taking into account the
expression for $\phi_{\pi^-}$ quoted in appendix~\ref{appB}, and defining
``parallel'' and ``perpendicular'' pieces $H^\pm_{\parallel,L}$ and
$H^\pm_{\perp,L}$, one gets
\begin{alignat}{2}
H^\pm_{\parallel,L} \ &= \ H^0_{L} \pm H^3_{L} & \ &= \
-\sqrt{2} \left(f^{(A1)}_{\pi^-} \mp f^{(V)}_{\pi^-} \right)
\left( {\cal D}^0 \pm {\cal D}^3 \right) W^-_{\bar p}(x) \nonumber\\
&&&= \ i \sqrt{2}\ \left(f^{(A1)}_{\pi^-} \mp  f^{(V)}_{\pi^-} \right)
\left( E_{\pi^-} \pm p_z \right) \ W^-_{\bar p}(x)\ ,
\\[.3cm]
H^\pm_{\perp,L} \ &= \ H^1_{L} \pm i\, H^2_{L} & \ &= \
- \sqrt{2} \left(f^{(A1)}_{\pi^-} \pm f^{(A2)}_{\pi^-} - f^{(A3)}_{\pi^-}\right)
\left( {\cal D}^1 \pm i\, {\cal D}^2 \right) W^-_{\bar p}(x) \nonumber\\
&&&= \ \mp \sqrt{2}  \left(f^{(A1)}_{\pi^-} \pm f^{(A2)}_{\pi^-} - f^{(A3)}_{\pi^-}  \right)
\sqrt{( 2\ell + 1 \pm 1) B_e} \ W^-_{\bar p \pm 1}(x) \ ,
\end{alignat}
where we have used the notation $\bar p \pm 1 = (E_{\pi^-},\ell\pm
1,\imath,p_z)$.

Using these expressions together with the explicit form of the functions
$U^-_l(x,\breve {q}, \tau)$, $V_{\nu_l}(x,\breve k, R)$ and $W^-_{\bar
p}(x)$, one can perform the spatial integral in eq.~\eqref{transition} to
get
\begin{eqnarray}
\langle\, l^-(\breve q , \tau)\, \bar \nu_l(\breve k, R ) | \mathcal{L}_W |
\pi^-(\breve p)\,\rangle &=& (2\pi)^3 \, \delta(E_{\pi^-} - E_l - E_{\bar\nu_l})\, \delta(p_z - q_z -
k_z)\,\times
\nonumber \\
& & \delta_{\ell-\imath, n-\upsilon + \jmath - 1}\, {\cal M}(\breve p, \breve q, \breve k, \tau)\ .
\label{amp}
\end{eqnarray}
The explicit form of the function ${\cal M}(\breve p, \breve q, \breve k,
\tau)$, as well as details of the calculation, can be found in
ref.~\cite{Coppola:2019wvh}. As expected from the symmetries of the
Lagrangian, eq.~\eqref{amp} shows the conservation of the total energy and
the $z$ component of the momentum. Moreover, from table~\ref{Tprop} it is
seen that the Kronecker delta in eq.~(\ref{amp}) implies
$j_z^{(\pi^-)}=j_z^{(l^-\,)}+j_z^{(\bar \nu_l)}$, i.e., the $z$ component of
the total canonical angular momentum is also conserved. This is not a
general property but a particular feature of the calculation in the
symmetric gauge, in which the Lagrangian is invariant under axial rotations.
We recall that, in the presence of the external magnetic field, the
canonical angular momentum is not a gauge invariant quantity and does not
represent a physical observable.

\subsection{Partial decay width}
\label{subsec:decayW}

The width for the $\pi^-\to l^-\,\bar\nu_l$ decay is given by
\begin{equation}
\Gamma_l^-\,(B)\ = \ \lim_{L,\,T\rightarrow\infty}  \sum_{\tau=1,2} \,\sum_{n,\upsilon,\jmath}
\int \!\frac{dq_z}{(2\pi)^3 2E_l} \dfrac{dk_z \ dk_\perp \ k_\perp}{(2\pi)^2 2E_{\bar\nu_l}}
\frac{|\langle\, l^-\,(\breve q ,\tau)\, \bar \nu_l( \breve k, R)
| \mathcal{L}_W | \pi^-(\breve p)\,\rangle |^2}{2 (2\pi)^2 E_{\pi^-} L \, T}\ , \label{uno}
\end{equation}
where $T$ and $L$ are the time interval and length on the $z$-axis in which
the interaction is active. At the end of the calculation, the limit $L,T\to
\infty$ should be taken. From the result in eq.~(\ref{amp}) we get
\begin{align}
\Gamma_l^-\,(B)\, \ = \ & \frac{1}{16\pi E_{\pi^-}}
\sum_{n,\upsilon=0}^\infty \sum_{\jmath=-\infty}^\infty\,
\int \frac{dq_z \ dk_z \ dk_\perp \ k_\perp}{(2\pi)^2 E_l \,E_{\bar\nu_l}}
\, \times \nonumber \\
& \delta(E_{\pi^-} - E_l - E_{\bar\nu_l}) \ \delta(p_z - q_z - k_z) \
\delta_{\ell-\imath, n-\upsilon-1 + \jmath} \
\overline{\big|{\cal M}_{\pi^-\to\, l^-\,\bar\nu_l}\big|^2}\ ,
\label{gamgen}
\end{align}
where
\begin{equation}
\overline{\big|{\cal M}_{\pi^-\to\, l^-\,\bar\nu_l}\big|^2} \ = \
\sum_{\tau=1,2}\Big|{\cal M}(\breve p, \breve q, \breve k, \tau)\Big|^2 \ .
\label{ampa}
\end{equation}

Now, as it is usually done, we concentrate on the situation in which the
decaying pion is in the lowest energy state. This corresponds to $\ell =0$
and $p_z =0$, hence $E_{\pi^-} = (m_{\pi^-}^2+B_e)^{1/2}$. Here we will
quote the final expression obtained for the decay width. Details of the
calculation can be found in ref.~\cite{Coppola:2019wvh}. The result can be
expressed in terms of three form factor combinations $a_{\pi^-}$,
$b_{\pi^-}$ and $c_{\pi^-}$, given by
\begin{equation}
a_{\pi^-} \ = \ f^{(A1)}_{\pi^-} - f^{(V)}_{\pi^-}\ , \qquad b_{\pi^-} \ = \ f^{(A1)}_{\pi^-} + f^{(V)}_{\pi^-}\ ,
\qquad c_{\pi^-} \ = \ f^{(A1)}_{\pi^-} + f^{(A2)}_{\pi^-} - f^{(A3)}_{\pi^-}\ .
\label{defabc}
\end{equation}
One has~\cite{Coppola:2019wvh}
\begin{equation}
\Gamma_l^-\,(B) \ = \
\dfrac{G_F^2\cos^2\theta_c}{2\pi\,E_{\pi^-}^2}\, B_e\,\sum_{n=0}^{n_{\rm
max}} \int_0^{u_{\rm max}}du\  \dfrac{1}{\rule{0cm}{0.4cm}\bar{k_z}(u)}\;\dfrac{u^{n-1}}{n!} \;e^{-u}\,
A^{(n)}_{\pi^-}(u)\ ,
\label{gamgenfinal}
\end{equation}
where the function $A^{(n)}_{\pi^-}(u)$ is given by
\begin{eqnarray}
\hspace{-0.2cm} A^{(n)}_{\pi^-}(u) & = & \big[E_{\pi^-}^2 -2B_e(n-u)-m_l^2\big] \, \times  \nonumber \\
& & \left[\frac{m_l^2}{2}\,(n |a_{\pi^-}|^2 + u |b_{\pi^-}|^2) + B_e (n-u)
(n|a_{\pi^-} - c_{\pi^-} |^2+u|b_{\pi^-} - c_{\pi^-}|^2)\right] +
\nonumber \\
& & 2B_e u
\big[E_{\pi^-}^2(n|a_{\pi^-} - b_{\pi^-} |^2-(n-u)|b_{\pi^-} - c_{\pi^-} |^2)+(n-u)\,m_l^2|c_{\pi^-} |^2\,\big] \ ,
\label{afull}
\end{eqnarray}
and we have used the definitions $u_{\rm max} =
\big(E_{\pi^-}-\sqrt{2nB_e+m_l^2}\big)^2/\,(2B_e)$, $n_{\rm max} \ = (E_{\pi^-}^2-m_{l}^{2})/(2B_{e})$ and
\begin{equation}
\bar{k_z}(u) \ = \ \dfrac{1}{2 E_{\pi^-}}
\left\{\Big[E_{\pi^-}^2-2B_e(n-u)-m_l^2\Big]^2
-8B_e\,E_{\pi^-}^2\,u\right\}^{1/2}\ .
\end{equation}
The integration variable chosen here is $u=k_\perp^2/(2B_e)$. The sum over
$\jmath$ and the integrals over $q_z$ and $k_z$ can be calculated with the
help of the deltas, while the sum over $\upsilon$ can be performed
analytically.

As expected, the decay width does not depend on the quantum number $\imath$.
The latter determines the canonical angular momentum $j_z^{(\pi^-)}$ of the
decaying pion, which, as stated, is a gauge dependent quantity. Though the
expression of the decay amplitude will vary in general for different gauge
choices, it is clear that the result for the decay width in
eq.~(\ref{gamgenfinal}) has to be gauge independent. Indeed, the same result
for $\Gamma_l^-\,(B)$ has been found in ref.~\cite{Coppola:2018ygv} using
the Landau gauge.

As discussed in the previous subsection, the decay constants in
eq.~(\ref{defabc}) parameterize the most general form of the pion-to-vacuum
vector and axial vector hadronic matrix elements. Their theoretical
determination would require either to use LQCD simulations or to rely on
some hadronic effective model. Before addressing possible estimates for
these quantities, let us analyze how ``non-helicity suppression'' is
realized in eq.~(\ref{gamgenfinal}). Once again we concentrate in the case
of a large external magnetic field. Since the pion is built of charged
quarks, the pion mass will depend in general on the magnetic field. Now, if
the mass growth is relatively mild, for large magnetic fields one should get
$B_e > m_{\pi^-}^2-m_l^2$. In fact, this is what one obtains from lattice QCD
calculations~\cite{Bali:2018sey} as well as from effective approaches like
the Nambu-Jona--Lasinio model~\cite{Coppola:2018vkw}, for values of $B_e$
say $\gtrsim 0.05$~GeV$^2$. According to the above expressions, this
implies $n_{\rm max} = 0$, hence the outgoing muon or electron (let us
assume that the energy is below the $\tau$ production threshold) is expected
to lie in its LLL ($n=0$), where only one polarization state is allowed. A
further simplification can be obtained when the squared lepton mass can be
neglected in comparison with $B_{e}$ (or, equivalently, in comparison with
$E_{\pi^-}^2$, which is expected to grow approximately as $B_{e}$). For
$m_l\ll B_e$, one can take $m_l\to 0$. Then, $E_l = \bar k_z$ and the
integral over $k_\perp$ extends up to $E_{\pi^-}$. In this limit the decay
width is given by
\begin{equation}
\Gamma_l^-(B)\Big|_{\substack{n_{\rm max}=0\\\;\;\;\,m_l=0}} \ = \
\frac{G_{F}^{2}\cos^{2}\theta_{c}}{\pi}\;\frac{B_{e}^{2}}{E_{\pi^-}}
\Big[1-\Big(1+\frac{E_{\pi^-}^{2}}{2B_{e}}\Big)\,
e^{-E_{\pi^-}^{2}/(2B_{e})}\,\Big]
\left|\, f^{(V)}_{\pi^-}- f_{\pi^-}^{(A2)}+f_{\pi^-}^{(A3)}\right|^{2} \ .
\label{largeb}
\end{equation}
As anticipated, there is no helicity suppression, and the width does not
vanish in the $m_l=0$ limit. In fact, it turns out to grow with the magnetic
field as $B_{e}^{2}/E_{\pi^-}\,$, with some suppression due to the factor in
square brackets. Clearly, the physical relevance of eq.~(\ref{largeb})
depends on whether the term proportional to the form factor combination on
the right hand side is the dominant one in the full expression for the decay
width. As can be seen from eq.~(\ref{afull}), the terms involving the form
factor $f^{(A1)}_{\pi^-}$ ---which, in general, would compete with the form
factors in eq.~(\ref{largeb})--- become negligible in the limit $m_l\to 0$.
While in the case of the $\pi^-$ decay to $e^-\bar \nu_e$ this should be a
good approximation already for $B_e\sim 0.05$~GeV$^2$, for decays into muons
(and taus) the situation is less clear, and corrections arising
from a nonzero lepton mass should be taken into account.

\section{Numerical results within the NJL model}
\label{sec:results}

In order to provide actual estimates for the magnetic field dependence of
the $\pi^-$ decay width we need some input values for the decay constants.
Although some results have been provided by existing LQCD
simulations~\cite{Bali:2018sey}, present lattice analyses involve relatively
large error bars and, moreover, only include the calculation of the form
factors $f_{\pi^-}^{(A1)}$ and $f_{\pi^-}^{(V)}$. Therefore, we will
consider here the values calculated in ref.~\cite{Coppola:2019uyr} for all
four form factors in the framework of the NJL model. In fact, beyond the
first lattice data points, results from ref.~\cite{Bali:2018sey} show an
overall increase in $f_{\pi^-}^{(A1)}$ with the magnetic field, in
qualitative agreement with the values obtained from NJL model
calculations~\cite{Coppola:2019uyr}. For $f_{\pi^-}^{(V)}$, NJL predictions
are compatible within errors with lattice data, which have been obtained for
$eB$ up to 0.3~GeV$^2$~\cite{Bali:2018sey,Coppola:2019uyr}.

\subsection{\texorpdfstring{$\bm{\Gamma_e^-}\:$}{} and \texorpdfstring{$\bm{\Gamma_\mu^-}\:$}{} decay widths}

\begin{figure}[htb]
\centering{\includegraphics[width=1.05\textwidth]{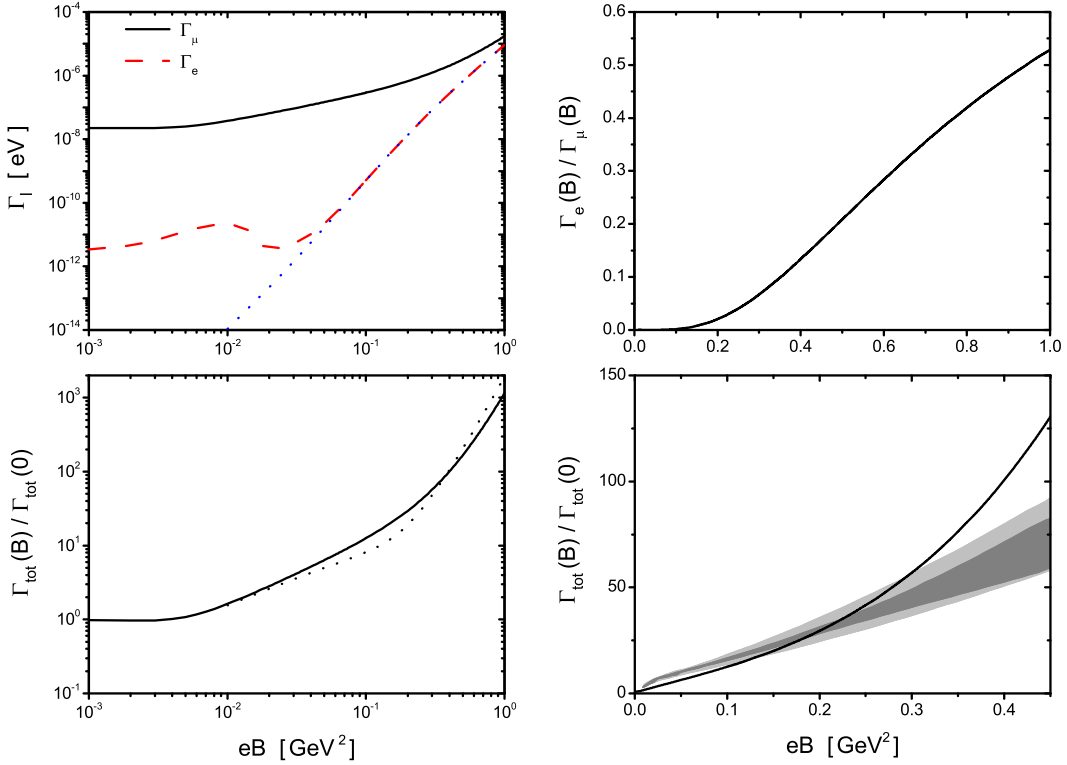}} \caption{(Color
online) Upper left panel: $\pi^-$ partial decay widths into $e^-\bar\nu_e$
(dashed line) and $\mu^-\bar\nu_\mu$ (full line), and $n=0$ asymptotic
contribution for $m_l=0$ (dotted line) as functions of $eB$. Upper right
panel:  ratio $\Gamma_e/\Gamma_\mu$ as a function of $eB$. Lower panels:
total decay width as a function of $eB$, normalized to its value at $B=0$,
shown in logarithmic scale (left) and linear scale (right). In the lower
left panel, the dotted line corresponds to the normalized total width in the
absence of the vector channel (i.e., taking $f_{\pi^-}^{(V)} = 0$). LQCD
bands quoted in ref.~\cite{Bali:2018sey} (see text) are included in the
lower right panel for comparison. Our results correspond to the model in
ref.~\cite{Coppola:2019uyr}, parameter Set I.} \label{Fig_Decay}
\end{figure}

Our results for the $\pi^-$ decay widths are shown in
figure~\ref{Fig_Decay}. They correspond to the parameter set denoted by
``Set I'' in ref.~\cite{Coppola:2019uyr}. In the upper left panel we quote
the $\pi^-$ partial decay widths to both $\mu^-\bar \nu_\mu$ and $e^-\bar
\nu_e$ as functions of $eB$, in a logarithmic scale. It is seen that the
partial widths become strongly enhanced when the magnetic field is increased
above say 0.1~GeV$^2/e$. This enhancement is more pronounced for the decay
to $e^-\bar \nu_e$ (dashed line), since for low values of $B$ helicity
suppression becomes important. The bump observed in this curve for $eB\sim
10^{-2}$~GeV$^2$ is due to the fact that this region is dominated by the
$n=1$ Landau level contribution, which disappears at about $eB\sim 2\times
10^{-2}$~GeV$^2$ leaving $n=0$ as the only energetically allowed electron
Landau level. The dotted line in the graph corresponds to the asymptotic
decay width quoted in eq.~(\ref{largeb}). In the upper right panel we quote
the ratio $\Gamma_e/\Gamma_\mu$ as a function of $eB$. The absence of
helicity suppression leads to a strong increase of this ratio with the
magnetic field, reaching a value of about 0.5 for $eB\simeq 1$~GeV$^2$,
while for $B=0$ one has $\Gamma_e/\Gamma_\mu\simeq 1.2\times 10^{-4}$. In
the lower panels we show the behavior of the total decay width $\Gamma_e +
\Gamma_\mu$, normalized to its value at $B = 0$. For this effective model
the enhancement factor is found to be about 1000 for $eB\simeq 1$~GeV$^2$.
Left and right panels show our results in logarithmic and linear scales,
respectively. To have an estimation of the relative significance of
the contribution coming from the vector piece of the hadronic amplitude, in
the left panel we show with a dotted line the result obtained for the total
width after setting $f_{\pi^-}^{(V)} = 0$. For large $B$ the correction will
be given by a global factor, as can be seen from eq.~(\ref{largeb}). In the
right panel we include for comparison the results arising form LQCD
calculations quoted in ref.~\cite{Bali:2018sey}, which cover values of $eB$
up to about 0.45~GeV$^2$. Dark and light gray regions correspond to
staggered and quenched Wilson quarks, respectively. Although these LQCD
results also predict a significant growth of the total width with the
magnetic field, it is seen that in our case the slope of the curve gets more
rapidly enhanced with $B$. This is, in part, due to the $e^-\bar\nu_e$
channel contribution. It is worth to remark that our results for the ratio
$\Gamma_e/\Gamma_\mu$ are different from those obtained in
ref.~\cite{Bali:2018sey}, where helicity suppression leads to a ratio of the
order of $10^{-5}$ that becomes almost independent of the magnetic field.
Finally, it is important to mention that the results in
figure~\ref{Fig_Decay} do not depend significantly on the model
parametrization (e.g.~it is seen that the results for parameter Sets II and
III of ref.~\cite{Coppola:2019uyr} do not differ from those in
figure~\ref{Fig_Decay} by more than 3\%).

\subsection{Angular distribution of outgoing neutrinos}

It is also interesting to discuss with some detail the angular distribution
of the outgoing antineutrinos. While for $B=0$ the distribution
is isotropic, this changes significantly in the presence of a large
magnetic field. Denoting $w=\cos\theta =k_z/|\vec k|$, the differential
decay rate can be written as
\begin{equation}
\dfrac{d\Gamma_l^-\,(B)}{dw} \ = \ \dfrac{G_F^2\cos^2\theta_c}{4\pi} \,\sum_{n=0}^{n_{\rm max}}\,
\dfrac{(1-r)^2}{r\,(1-w^2)^2} \, \dfrac{u^{n-1}}{n!} \;e^{-u}\,
\left[ |w|\dfrac{A^{(n)}_{\pi^-}(u)}{\rule{0cm}{0.4cm}\bar{k_z}(u)}+w\, B^{(n)}_{\pi^-}(u) \right]\ ,
\label{differential}
\end{equation}
where
\begin{equation}
r \ = \ \dfrac{1}{E_{\pi^-}}\sqrt{E_{\pi^-}^2-\left(E_{\pi^-}^2-2nB_e-m_l^2 \right)(1-w^2)}\ ,
\qquad
u \ = \ \dfrac{E_{\pi^-}^2}{2B_e}\, \dfrac{(1-r)^2}{(1-w^2)}\ ,
\end{equation}
and the function $B^{(n)}_{\pi^-}(u)$ is defined as
\begin{equation}
B^{(n)}_{\pi^-}(u) \ = \ E_{\pi^-} \! \left[ \left(u |b_{\pi^-}|^2 - n |a_{\pi^-}|^2 \right) m_l^2
+2B_e(n-u)\left( u|b_{\pi^-} - c_{\pi^-}|^2 - n|a_{\pi^-} - c_{\pi^-} |^2 \right) \right]\
.
\end{equation}
The term proportional to $B^{(n)}_{\pi^-}(u)$ in eq.~(\ref{differential}) vanishes after
integration over $w$, therefore it does not contribute to the total decay width.

\begin{figure}[htb]
    \centering{}\includegraphics[width=0.9\textwidth]{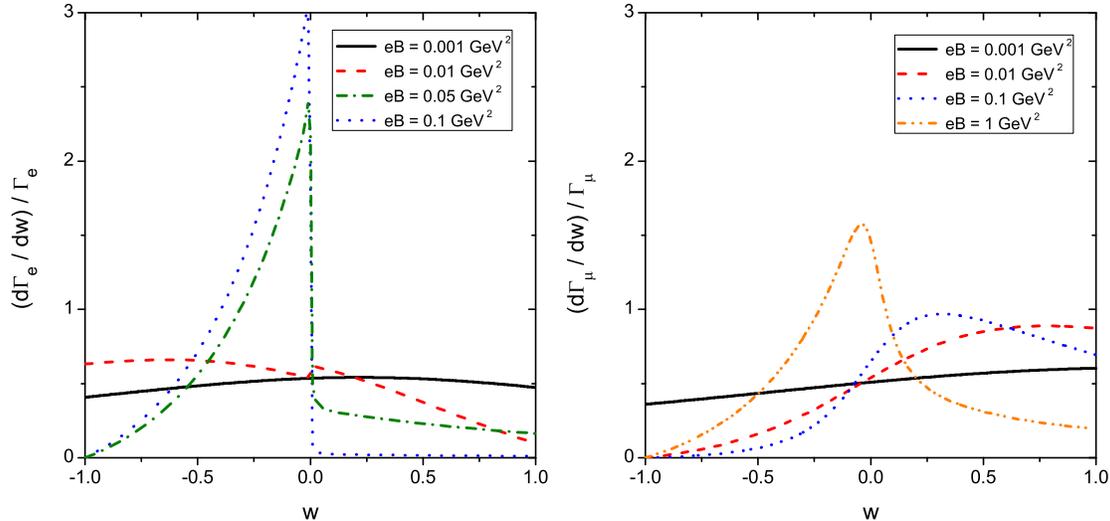}
\caption{(Color online) Normalized differential partial decay widths of the $\pi^-$
into $e^-\bar{\nu}_e$ (left) and $\mu^-\bar{\nu}_\mu$ (right), as functions of $w=\cos
\theta$ for selected values of $eB$. The results correspond to the model in
ref.~\cite{Coppola:2019uyr}, parameter Set I.}
\label{Fig_Angular}
\end{figure}

Once again, to get definite predictions for the angular
distributions we rely on the values for the pion mass and decay
constants obtained in ref.~\cite{Coppola:2019uyr} within the NJL
model, taking the parameter Set I. Our numerical results for the
normalized differential partial decay widths are shown in
figure~\ref{Fig_Angular}, where several representative values of
$eB$ are considered. Left and right panels correspond to $\pi^-$
decays into $e^-\bar{\nu}_e$ and $\mu^-\bar{\nu}_\mu$,
respectively. It is seen that the fraction of antineutrinos that
come out in the half-space $w > 0$ fluctuates when the
magnetic field is increased, becoming strongly suppressed for
values of $eB$ much larger than the lepton mass squared.
This can be qualitatively understood as follows. When $eB \gg m_l^2$, only $n=0$ is allowed. In addition, in the massless limit the lepton has to be left-handed, therefore from eq.~\eqref{chirality} one gets $q_z > 0$. Conservation of the $z$ component of total momentum implies $q_z+k_z = p_z =0$. Hence, for large $B$, in the $m_l \to 0$ limit all antineutrinos should be produced with momenta in the half-space $k_z < 0$. Indeed, for $m_l=0$ and $n=0$ the normalized  differential decay width is given by
\begin{equation}
\frac{1}{\Gamma_l^-\,(B)}\;\dfrac{d\Gamma_l^-\,(B)}{dw} \ = \ \left\{
\begin{array}{ll}
2\lambda^2\dfrac{(1+w)}{(1-w)^3} \, \dfrac{e^{-\lambda(1+w)/(1-w)}}
{1-(1+\lambda)\,e^{-\lambda}} & \ \ {\rm if}\ \ w \leq 0 \\
\rule{0cm}{0.8cm} 0 & \ \ {\rm if}\ \ w > 0
\end{array}
\right. \ \ ,
\label{ml0}
\end{equation}
where $\lambda = E_{\pi^-}^2/(2B_e)$.
In addition, it is worth
noticing that for large values of $B$ most antineutrinos come out
with low $|k_z|$, i.e.\ in directions approximately perpendicular
to the magnetic field.

\section{Summary and conclusions}
\label{sec:summary}

In this article we get an estimation of the effect of an external
uniform magnetic field on the magnitude of the decay rate $\Gamma(\pi^-\to
l^-\,\bar{\nu}_l)$ and the angular distribution of the antineutrinos in the
final state. Our analysis takes into account the contribution of all four
possible $\pi^-$ decay form factors. The values of these constants and that
of the pion mass are taken from a NJL model for effective strong
interactions, considering the $\pi^-$ in its lowest possible energy state.
Our results show that the total decay rate $\Gamma_e+\Gamma_\mu$
becomes strongly increased with respect to its value at $B=0$, the
enhancement factor ranging from $\sim 10$ for $eB=0.1$~GeV$^2$ up to $\sim
10^3$ for $eB=1$~GeV$^2$. Moreover, owing to the presence of the new decay
constants and the features of nonzero $B$ kinematics, it is found that the
decay width $\Gamma_l^-$ does not vanish in the limit $m_l=0$. As a
consequence, for large values of $B$ the ratio $\Gamma_e/\Gamma_\mu$
changes dramatically with respect to the $B=0$ value (of about $1.2\times
10^{-4}$), reaching a magnitude of $\sim 0.5$ at $eB\simeq 1$~GeV$^2$. This
could be interesting e.g.~regarding the expected flavor composition of
neutrino fluxes coming from the cores of magnetars and other stellar
objects. Finally, it is found that for large $B$ the angular distribution of
outgoing antineutrinos is expected to be highly anisotropic, showing a
significant suppression in the direction of the external field.

\acknowledgments

This work has been supported in part by Consejo Nacional de Investigaciones
Cient\'ificas y T\'ecnicas and Agencia Nacional de Promoci\'on Cient\'ifica
y Tecnol\'ogica (Argentina), under Grants No.~PIP17-700 and No.
PICT17-03-0571, respectively; by the National University of La Plata
(Argentina), Project No.~X824; by the MICINN (Spain), under Contract
No.~FPA2016-77177-C2-1-P, PID2019-105439GB-C21 and by EU Horizon 2020 Grant
No.~824093 (STRONG-2020).

\appendix
\section{Axial rotations and gauge invariance}
\label{appA}

The consequences of the invariance of the physical system under rotations in
the plane perpendicular to the magnetic field, as well as the relation of
this invariance with the conservation of the corresponding component of the
angular momentum, are delicate issues that deserve some extra comments.

Let us consider a charged pion in the presence of a uniform magnetic field,
using the conventions stated in the main text of this work. Given the
symmetry of the physical system, any observable is expected to be invariant
under rotations about the $z$ axis. However, it is worth noticing that the
Lagrangian and the action that describe the system at the quantum mechanical
level are given in terms of the electromagnetic four-vector potential. Thus,
they are not necessarily invariant under these rotations. For the particular
case of the symmetric gauge used in this work, rotational symmetry is
manifest. However, in general this will be not true for other gauges. To
illustrate this point let us consider the Landau gauge (LG), in which
$A^{\mu}=\left(0,\,0,\,x\,B,\,0\right)$. A spatial rotation by an angle
$\theta$ about the $z$ axis changes $A^{\mu}$ into $A^{\prime\mu}$, which is
given by
\begin{equation}
A^{\prime\mu}=\left(0,\,-\sin\theta\left(x\cos\theta+y\sin\theta\right)\,B,
\,\cos\theta\left(x\cos\theta+y\sin\theta\right)\,B,\,0\right)\ .\label{1}
\end{equation}
The breakdown of the invariance of the Lagrangian under this rotation is
manifest. Moreover, it can be checked that in the LG neither the $z$
component of the canonical angular momentum nor that of the mechanical
angular momentum commute with the Hamiltonian of the system, i.e., they are
not conserved quantities. In order to reconcile this result with the
expected invariance of the physical quantities under spatial rotations, we
can observe that $A^{\prime\mu}$ can also be written as
\begin{equation}
A^{\prime\mu}=\left(0,\,0,\,x\,B,\,0\right)+\partial^{\mu}\chi\ ,\label{2}
\end{equation}
with
\begin{equation}
\chi\ = \ \frac{B \sin\theta}{2}\left(x^{2}\,\cos\theta+2x\,y\,\sin\theta-
y^{2}\,\cos\theta\right)\ .
\label{3}
\end{equation}
In this way, it is seen that the rotated system is connected to a gauge
transformed system through a gauge transformation defined by $\chi$. This
shows that, in the Landau gauge, performing a spatial rotation about the $z$
axis is equivalent to performing a specific gauge transformation. Thus, in
this gauge the expected invariance of physical observables under spatial
rotations is guaranteed by the gauge invariance of the system.

Let us illustrate the previous statement for the case of the $\pi^-$ field.
As discussed in appendix A.2 of ref.~\cite{Coppola:2018ygv}, in the Landau gauge
the $\pi^-$ wavefunction can be written as
\begin{equation}
\mathbb{F}_{\tilde p}(x) \ = \ \tilde N_{\ell}\,e^{-iE_{\pi^-}t}\,e^{ip_{y}y}\,e^{ip_{z}z}\,
D_{\ell}\Big(\sqrt{2B_e}\,\big(x+\frac{p_{y}}{B_e}\big)\Big)
\label{4}
\end{equation}
where $D_{\ell}(x)$ are cylindrical parabolic functions, and we have defined
$\tilde p =(E_{\pi^-},\ell,p_{y},p_{z})$ and $\tilde N_{\ell}=(4\pi B_e)^{1/4}
/\sqrt{\ell!}\,$. After a rotation by an angle $\theta$ about the $z$
axis, one gets a rotated wavefunction $\mathbb{F}_{\tilde p}^{\text{R}}(x)$
given by
\begin{equation}
\mathbb{F}_{\tilde p}^{\text{R}}(x)\ =\ \tilde N_{\ell}\,e^{-iE_{\pi^-}t}\,
e^{ip_{y}(-x\,\sin\theta+y\,\cos\theta)}\,e^{ip_{z}z}\,
D_{\ell}\Big(\sqrt{2B_{e}}\big(x\,\cos\theta+y\,\sin\theta+
\frac{p_{y}}{B_e}\big)\Big)\ .
\end{equation}
On the other hand, performing the gauge transformation defined in
eqs.~(\ref{2}) and (\ref{3}), the pion wave function in eq.~(\ref{4})
transforms into $\mathbb{F}_{\tilde p}^{\chi}\left(x\right)$, given by
\begin{equation}
\mathbb{F}_{\tilde p}^{\chi}(x)\ =\ e^{ie\chi}\,\mathbb{F}_{\tilde p} (x)
\ =\ e^{i\sin\theta\,(x^2\,\cos\theta+2
x\,y\,\sin\theta-y^2\cos\theta)B_e/2}\,\mathbb{F}_{\tilde
p}(x)\ .
\end{equation}
Obviously, $\mathbb{F}_{s,\bar{p}}^{\text{R}}\left(x\right)$ and
$\mathbb{F}_{s,\bar{p}}^{\chi}\left(x\right)$ are different. However, they
are connected in the sense that they share the same subspace of defined
values of energy and momentum $p_{z}\,$. This subspace is built varying the
value of $p_{y}\,$, which is not gauge invariant and therefore cannot be
taken as a physical quantity~\cite{Coppola:2018ygv}.

The fact that the functions $\mathbb{F}_{\tilde p}^{\text{R}}(x)$ and
$\mathbb{F}_{\tilde p}^{\chi}(x)$ belong to the same subspace of energy and
momentum $p_{z}\,$ can be verified by projecting one function onto the
other. One has
\begin{eqnarray}
\int d^{3}x\;\mathbb{F}_{\tilde p^\prime}^{\chi}(x)^\ast\;\mathbb{F}_{\tilde p}^{\text{R}}(x) 
& = &
(2\pi)^{2}\delta\left(p_{z}^{\prime}-p_{z}\right)\,\delta_{\ell \ell^\prime}\, \times \nonumber\\
&& \sqrt{\frac{2\pi}{B_e\sin\theta}}\,
e^{-i[(p_{y}^{2}+{p_y^\prime}^2)\cos\theta -2p_{y}p_y^\prime]/(2B_e\sin\theta)}
\,e^{-i[(\ell+1/2)\theta-\pi/4]}\ ,
\label{KG_L2_Rotacion}
\end{eqnarray}
which proves our statement, taking into account the (gauge independent)
relation $E_{\pi^-} = \sqrt{m_{\pi^-}^2+ (2\ell+1) B_e + p_z^2}$. As a check of
the completeness of the transformed functions, it can be seen that\\
\begin{equation}
\int \dfrac{dp_{z}^{\prime}\,dp_{y}^{\prime}}{(2\pi)^{3}} \,\sum_{\ell^{\prime}=0}^\infty\,
\int d^3x'\;\mathbb{F}_{{\tilde p}^{\prime\prime}}^{\text{R}}(x')^\ast\,
\mathbb{F}_{{\tilde p}^{\prime}}^{\chi}(x')\,
\int d^3x\;\mathbb{F}_{{\tilde p}^{\prime}}^{\chi}(x)^\ast\,
\mathbb{F}_{\tilde p}^{\text R}(x)  = 
(2\pi)^3\,\delta(p_{z}^{\prime\prime}-p_{z})\,\delta(p_{y}^{\prime\prime}-p_{y})\,
\delta_{\ell\ell^{\prime\prime}}\ .
\end{equation}

\section{Particle fields under a uniform magnetic field in the symmetric gauge}
\label{appB}

For convenience, we quote in this appendix the main
expressions for $\pi^-$, $l^-$ and $\nu_l$ fields in the presence
of a magnetic field together with the eigenvalues
of some relevant operators. For a more detailed description, see
e.g.~refs.~\cite{Coppola:2019wvh} and~\cite{Sokolov:1986nk}.

According to our conventions, the $\pi^-$ field can be written as~\cite{Coppola:2019wvh}
\begin{equation}
\phi_{\pi^{-}}(x) \ = \
\sum_{\ell,\imath=0}^\infty \int \frac{ dp_z}{(2\pi)^3\, 2 E_{\pi^-}}
\left[
a^-(\breve p)  \; W^-_{\bar{p}}(x) + a^{+}(\breve p)^\dagger  \; W^+_{\bar{p}}(x)^\ast \right]\ ,
\label{chargepionexp}
\end{equation}
where $\bar p = (E_{\pi^-},\breve{p})$, with $\breve{p} =
(\ell,\imath,p_z)$ and $E_{\pi^-} = \sqrt{m_{\pi^-}^2+ (2\ell+1) B_e + p_z^2}$. The functions $W^\pm_{\bar{p}}(x)$ are solutions of the
eigenvalue equation
\begin{equation}
{\cal D}_\mu {\cal D}^\mu \  W^\pm_{\bar{p}}(x) \ =
\ -  \left[ E_{\pi^-}^2 - (2 \ell+1) B_e - p_z^2 \right]  W^\pm_{\bar{p}}(x)\ ,
\label{ecautovbo}
\end{equation}
where ${\cal D}^\mu = \partial^\mu - i e A^\mu$. Using cylindrical
coordinates, their explicit form is given by
\begin{equation}
 W^\pm_{\bar{p}}(x) \ = \
\sqrt{2\pi} \ e^{-i(E_{\pi^-} t - p_z z)}\, e^{\mp i (\ell - \imath) \phi}
\, R_{\ell,\imath}(\rho)\ ,
\label{efes}
\end{equation}
where
\begin{equation}
R_{\ell,\imath}(\rho) \ = \ N_{\ell,\imath} \ \xi^{(\ell - \imath)/2} \  e^{-\xi/2} \ L_\imath^{\ell-\imath}(\xi)\ .
\end{equation}
Here we have used the definitions $N_{\ell, \imath} = (B_e \ \imath! /
\ell!)^{1/2}$ and $\xi = B_e \, \rho^2/2\,$, while $L_m^\alpha(x)$ are the
associated Laguerre polynomials.

The charged lepton fields in this gauge can be written as
\begin{equation}
\psi_l(x) \ = \
\sum_{\tau=1,2} \ \sum_{n,\upsilon=0}^\infty \int \! \frac{dq_z}{(2\pi)^3 \, 2 E_l}
\left[\,
b\left(\breve {q},\tau\right) \, U_l^-\left(x,\breve {q},\tau\right) + d\left(\breve {q},\tau\right)^\dagger
\,  V_l^{+}\left(x,\breve {q},\tau\right)\,
\right]\ ,
\label{fermionfieldBpart}
\end{equation}
where $\breve {q} = (n,\upsilon,q_z)$ and $E_l = \sqrt{m_l^2+ 2n B_e + q_z^2}$. For $n>0$, in the Weyl basis, the spinors in
eq.~(\ref{fermionfieldBpart}) are given by
\begin{eqnarray}
U_l^-\left(x,\breve{q},\tau \right) & = &
\frac{\sqrt{\pi}}{\sqrt{E_l + m_l}} e^{-i (E_l t - q_z z)} e^{i (n-\upsilon-1/2) \phi}\, \times \nonumber \\
&& \hspace{-.6cm}\left[
\delta_{\tau,1} \!
\left( \!\!\! \begin{array}{c}
 e^{-i\phi/2} \ \varepsilon_- \  R_{n-1,\upsilon}(\rho) \\
- i e^{i\phi/2} \ \sqrt{2 n B_e} \  R_{n,\upsilon}(\rho) \\
 e^{-i\phi/2} \ \varepsilon_+ \ R_{n-1,\upsilon}(\rho)\\
i e^{i\phi/2} \ \sqrt{2 n B_e} \ R_{n,\upsilon}(\rho) \\
\end{array}
\!\! \right)
+
\delta_{\tau,2}
\left( \!\!\! \begin{array}{c}
i e^{-i\phi/2} \ \sqrt{2 n B_e} \  R_{n-1,\upsilon}(\rho) \\
e^{i\phi/2} \ \varepsilon_+    \ R_{n,\upsilon}(\rho) \\
- i e^{-i\phi/2} \ \sqrt{2 n B_e} \  R_{n-1,\upsilon}(\rho) \\
 e^{i\phi/2} \ \varepsilon_-   \ R_{n,\upsilon}(\rho)\\
\end{array}
\!\! \right)
\right]\, ,
\end{eqnarray}
\begin{eqnarray}
V_l^+\left(x,\breve{q},\tau\right) & = &
\frac{\sqrt{\pi}}{\sqrt{E_l + m_l}} e^{i (E_l t - q_z z)} e^{i (n-\upsilon-1/2) \phi}\,\times \nonumber \\
&&\hspace{-.6cm} \left[
\delta_{\tau,1} \!
\left( \!\!\! \begin{array}{c}
- i e^{-i\phi/2} \ \sqrt{2 n B_e} \ R_{n-1,\upsilon}(\rho) \\
e^{i\phi/2} \ \varepsilon_+    \ R_{n,\upsilon}(\rho) \\
- i e^{-i\phi/2} \ \sqrt{2 n B_e} \ R_{n-1,\upsilon}(\rho) \\
 - e^{i\phi/2} \ \varepsilon_-   \ R_{n,\upsilon}(\rho)\\
\end{array}
\!\!\! \right) \! \!
+
\delta_{\tau,2}
\left( \!\! \begin{array}{c}
- e^{-i\phi/2} \ \varepsilon_- \  R_{n-1,\upsilon}(\rho) \\
- i e^{i\phi/2} \ \sqrt{2 n B_e} \ R_{n,\upsilon}(\rho) \\
 e^{-i\phi/2} \ \varepsilon_+ \ R_{n-1,\upsilon}(\rho)\\
- i e^{i\phi/2} \ \sqrt{2 n B_e} \ R_{n,\upsilon}(\rho) \\
\end{array}
\!\! \right)
\right]\, ,
\end{eqnarray}
where $\varepsilon_\pm = E_l + m_l \pm q_z$. In the particular case of the
lowest Landau level (LLL) $n=0$, from these equations it is seen that
$U_l^-(x,\breve{q},1) = V_l^+(x,\breve{q},2) = 0$, i.e., only one
polarization state is allowed in each case. Using the notation $\breve
q_{LLL} = (0,\upsilon,q_z)$, the explicit forms of the spinors are
\begin{align}
U_l^-\left(x,\breve q_{LLL}\right) &= \frac{\sqrt{\pi}}{\sqrt{E_l+m_l}} e^{-i (E_l t - q_z z)} e^{-i \upsilon \phi}
R_{0,\upsilon}(\rho) \left( \!\!\! \begin{array}{c}
0 \\
\ \varepsilon_+ \ \ \\
0 \\
\varepsilon_-\\
\end{array}
\!\!\! \right)\ ,
\label{ULLL}  \\
V_l^+\left(x,\breve q_{LLL}\right) &=
\frac{\sqrt{\pi}}{\sqrt{E_l+m_l}} e^{i (E_l t - q_z z)} e^{-i \upsilon \phi} R_{0,\upsilon}(\rho)
\left( \!\!\! \begin{array}{c}
0 \\
\varepsilon_+\\
0 \\
-\varepsilon_-\\
\end{array}
\!\!\! \right)\ . \label{VLLL}
\end{align}

It is interesting to consider in this context the canonical
orbital angular momentum operator $\vec l = \vec r \times \vec p$
and the spin operator $\vec S = \vec \Sigma/2$. Given the fact that the
magnetic field breaks rotational invariance, only the $z$
components of these operators are relevant. These are given by
$l_z = -i \partial/\partial\phi$ and $S_z = \mbox{diag}(1,-1,1,-1)/2$.
Defining the canonical total angular momentum as $j_z = l_z +
S_z$, one obtains
\begin{equation}
j_z |l(\breve q, \tau) \rangle \ = \ \left(n - \upsilon -\dfrac{1}{2}\right) \ |l(\breve q, \tau)
\rangle\ .
\end{equation}
Thus, as expected from axial symmetry, it is seen that for the charged
leptons in the symmetric gauge one can find energy eigenstates that are also
eigenstates of $j_z$. It is worth noticing that only the
total canonical angular momentum is well-defined, i.e., energy eigenstates
are not in general eigenstates of $l_z$ and $S_z$ separately.

Let us consider now the limit in which the charged lepton mass $m_l$
vanishes. This is interesting when the magnetic field is relatively strong,
say $B_e \gg m_l^2$. In the limit $m_l = 0$ the chirality operator
$\gamma_5$ becomes equivalent to the helicity operator and commutes with the
Hamiltonian. Consequently, one can obtain energy eigenstates of well defined
chirality/helicity as linear combinations of the two polarization states. In
the particular case of the LLL, since only one polarization state is
available, it has to be a helicity eigenstate. The corresponding particle
and antiparticle spinors are obtained from eqs.~(\ref{ULLL})
and~(\ref{VLLL}) taking $m_l =0$. It can be easily seen that in this
case the relations in eq.~(\ref{chirality}) are satisfied. In this way, for
large enough magnetic fields ---such that only the LLL is relevant and $m_l$
can be neglected--- a negatively charged lepton (like the muon or the
electron) is lefthanded if $q_z$ is positive, and it is righthanded
otherwise.

For the case of the $\pi^-$, from the above equations it is easy to see that
the canonical orbital angular momentum is given by
\begin{eqnarray}
l_z \ |\pi^-(\breve p) \rangle \ = \  (\ell - \imath) \, |\pi^-(\breve p)\rangle \ .
\end{eqnarray}
Since the $\pi^-$ is a spin zero particle, one has in this case $j_z = l_z\,$.

Finally, let us consider the neutrino and antineutrino fields. It is usual
to write these fields in terms of operators of well-defined linear momentum
$\vec k$. However, for our purposes it is convenient to expand the usual
plane wave functions in terms of eigenfunctions of $l_z$. Next, we couple
these wavefunctions to the eigenstates of $S_z$, and write the neutrino
and antineutrino states in terms of eigenstates of the total angular
momentum $j_z = l_z + S_z$. The resulting expansion for the fields
reads
\begin{equation}
\psi_{\nu_l}(x) \ = \
\sum_{\jmath=-\infty}^\infty \int \! \frac{dk_z}{2\pi}
\int_0^\infty\frac{dk_\perp \, k_\perp}{4\pi E_{\nu_l}}
\left[ b(\breve k, L) \ U_{\nu_l}(x,\breve k,L) + d(\breve k, R)^\dagger  \ V_{\nu_l}(x,\breve k,R)
\right]\ ,
\label{neutrinofield}
\end{equation}
where  $\breve k = (\jmath, k_\perp, k_z)$ and $E_{\nu_l} = E_{\bar\nu_l} =
\sqrt{k_\perp^2 + k_z^2}\,$. In the Weyl basis, the spinors $U_{\nu_l}$ and
$ V_{\nu_l}$ are given by
\begin{equation}
U_{\nu_l}\left(x,\breve k,L\right)\ = \
-\, i^{\jmath} \ e^{-i(E_{\bar\nu_l} t - k_z z)} \ e^{-i \jmath \; \phi}
\left( \!\!\! \begin{array}{c}
      \sqrt{ E_{\bar\nu_l} - k_z} \ J_{\jmath}(k_\perp \rho) \\
      i \sqrt{ E_{\bar\nu_l} + k_z}  \ e^{i \phi} \ J_{\jmath-1}(k_\perp \rho) \\ 0\\ 0
\end{array}
\!\!\! \right)\ ,
\label{Uspindos}
\end{equation}
\begin{equation}
V_{\nu_l}\left(x,\breve k,R\right)\ = \
- (-i)^{\jmath} \ e^{i(E_{\bar\nu_l} t - k_z z)} \ e^{-i \jmath \; \phi}
\left( \!\!\! \begin{array}{c}
      \sqrt{ E_{\bar\nu_l} - k_z} \ J_{\jmath}(k_\perp \rho) \\
   -i \sqrt{ E_{\bar\nu_l} + k_z} \ e^{i\phi} \ J_{\jmath-1}(k_\perp \rho)\\ 0\\ 0
\end{array}
\!\!\! \right)\ .
\end{equation}
Note that, as it is clear from the explicit form of the spinors, in the expansion
we have already taken into account that neutrinos (antineutrinos) are lefthanded (righthanded).

It is seen that antineutrino states satisfy
\begin{eqnarray}
j_z \, |\bar\nu_l(\breve k,R) \rangle & = &  \left(\jmath-\frac{1}{2}\right) \, |\bar\nu_l(\breve k,R)\rangle \
, \nonumber \\
\gamma_5 \, |\bar\nu_l(\breve k,R) \rangle & = & \hat k \cdot \vec \Sigma\, |\bar\nu_l(\breve k,R)\rangle
\ = \ |\bar\nu_l(\breve k,R)\rangle\ .
\end{eqnarray}

\end{document}